\documentclass{article}
\usepackage{silence}
\PassOptionsToPackage{numbers, compress}{natbib}
\usepackage[dblblindworkshop,preprint]{neurips_2026}
 
\usepackage[utf8]{inputenc} 
\usepackage[T1]{fontenc}    
\usepackage{hyperref}       
\usepackage{url}            
\usepackage{booktabs}       
\usepackage{amsfonts}       
\usepackage{nicefrac}       
\usepackage{microtype}      
\usepackage{xcolor}         

\usepackage{graphicx}
\usepackage{subcaption}
\usepackage{amssymb}
\usepackage{placeins}
\usepackage{amsmath}

\newif\ifanonymousartifacts
\anonymousartifactsfalse

\title{Learning the Geometry of Collider Events with Metric-Aware Deep Sets}

\author{%
  Lauren~Hay\\
  Brown University / IAIFI
  \And
  Rishabh~Jain\thanks{rishabh\_jain@brown.edu}\\
  Brown University
  \And
  Matt~LeBlanc\thanks{matt\_leblanc@brown.edu}\\
  Brown University / IAIFI
  \And
  Jennifer~Roloff \\
  Brown University
}

\begin{document}

\maketitle

\begin{abstract}

Optimal transport gives structured data a geometry, but exact evaluation is costly in large pairwise analyses that exploit relationships among distances. 
Learned surrogates are faster, but need not preserve this metric structure.
We develop a Deep Sets surrogate for OT between variable-size weighted point clouds that enforces non-negativity, exchange symmetry, and zero self-distance, leaving the triangle inequality unconstrained.
Applied to the Energy Mover’s Distance between collider events in a particle physics application, the Metric-Aware Particle Flow Network achieves percent-level mean absolute percentage error while significantly improving inference throughput over other exact and approximate methods surveyed.
The architectural constraints are found to improve properties that are not explicitly enforced: across \(10^6\) held-out event triplets, triangle-inequality violations fall from 199 for a matched unconstrained network to 2, and the maximum from 149.5 to 5.8 GeV.
These results demonstrate that targeted inductive biases can yield fast neural surrogates with substantially improved geometric fidelity.

\end{abstract}

\section{Introduction}

Optimal transport (OT) equips distributions and structured data with a useful geometry~\cite{peyre2019computational}, but its computational cost limits applications involving many complex objects.
A dataset of $n$ objects requires $O(n^2)$ pairwise comparisons, and the cost of each transport problem grows with the support sizes of the objects being compared.
This compounds in nearest-neighbor searches, dataset clustering \& embedding, and related workflows that depend on the relationships among many distances.

Learning the transport cost can amortize this computation, but it introduces two distinct approximation problems.
\emph{Numerical accuracy} asks whether each predicted value is close to the exact cost.
\emph{Geometric fidelity} asks whether the predictions collectively behave like a distance: whether they are non-negative and symmetric, vanish on the diagonal, and satisfy the triangle inequality.
Low regression error alone guarantees neither of these properties, so an accurate surrogate can still distort the geometry used by downstream algorithms.
We therefore ask how much of the target geometry can be restored through architectural constraints without requiring the full metric structure to be enforced.

Fast alternatives to exact OT include entropic and sliced approximations~\cite{feydy_sinkhorn_2019,bonneel_sliced_2015}.
In collider physics, complementary approaches have used linearized transport representations, learned metric embeddings, and neural EMD approximations~\cite{Cai:2020vzx,Tsan:2021brw,Park:2022zov,Kitouni:2022qyr}.
Our focus is a directly supervised pairwise regressor: rather than changing the target distance, we encode part of its geometry in a permutation-invariant architecture and measure both scalar and geometric error.

Collider physics data provides a demanding concrete test bed for this general problem: each collision is a variable-size weighted point cloud, often with hundreds of particles, and the Energy Mover's Distance (EMD) defines an unbalanced OT metric between such events~\cite{Komiske:2018cqr,Komiske:2020qhg}.
The EMD has enabled measurements at the CERN Large Hadron Collider~\cite{ATLAS-STDM-2020-20,CMS-SMP-23-008,Cesarotti:2020hwb,cesarotti_field_2025} and improvements to collider simulation~\cite{doherty2026}.
An exact comparison of two typical events may require optimizing over a roughly $200\times200$ particle-level cost matrix, making dense event-distance matrices expensive.
The same mathematical setting encompasses OT between other weighted point clouds and empirical distributions.

We introduce the Metric-Aware Particle Flow Network (MA-PFN), a Deep Sets surrogate for the EMD.
A shared point encoder and symmetrized joint head guarantee non-negativity, exchange symmetry, and zero self-distance.
The separation of distinct inputs and the triangle inequality remain unconstrained.
Against a matched unconstrained PFN, the MA-PFN achieves sub-percent mean absolute percentage error and a dense-pair throughput advantage over exact CPU solvers that grows with support size.
The imposed constraints also improve the untargeted axiom: across $10^6$ held-out triplets, triangle-inequality violations fall from 199 to 2, and the maximum from $149.5$ to $5.8$~GeV.
We additionally assess physically motivated low-energy and collinear deformations.
Together, these results show that lightweight inductive biases can preserve substantially more of a learned distance's geometry than the properties they enforce directly.

\section{Methodology}

\subsection{Data and target distance}\label{sec:data}

We represent a collider event as a variable-size weighted point cloud in detector coordinates,
\begin{equation}
    \begin{aligned}
        \mathcal E&=\left\{x_i=(p_{T,i},\eta_i,\phi_i)\right\}_{i=1}^{N},
        &\eta_i&=-\log\tan\left(\frac{\theta_i}{2}\right),\\
        \Delta R_{ij}&=\sqrt{(\eta_i-\eta'_j)^2+\Delta\phi_{ij}^{\,2}}.
    \end{aligned}
    \label{eq:event-representation}
\end{equation}
where $N$ is the multiplicity, $p_T$ is momentum transverse to the beam, $\theta$ is the polar angle, $\eta$ is pseudorapidity, $\phi$ is the azimuthal angle, $\Delta\phi_{ij}$ is the periodic azimuthal difference, and $\Delta R$ is the angular ground distance between two events.

The Energy Mover's Distance (EMD) is an unbalanced OT metric: it transports transverse momentum between two such events and penalizes unmatched weight,
\begin{equation}
    \mathrm{EMD}_{\beta,R}(\mathcal E,\mathcal G)
    =\min_{f_{ij}\geq0}
    \sum_{i=1}^{N}\sum_{j=1}^{M}f_{ij}
    \left(\frac{\Delta R_{ij}}{R}\right)^\beta
    +\left|\sum_i p_{T,i}-\sum_j p'_{T,j}\right|,
    \label{EMD}
\end{equation}
where $f_{ij}$ is the transverse momentum transported from particle $i$ to particle $j$, subject to
\begin{equation}
    \sum_j f_{ij}\leq p_{T,i},\qquad
    \sum_i f_{ij}\leq p'_{T,j},\qquad
    \sum_{ij}f_{ij}
    =\min\left(\sum_i p_{T,i},\sum_j p'_{T,j}\right).
    \label{eq:emd-constraints}
\end{equation}
The first term is the angular transport cost; the second penalizes unmatched total $p_T$, so the EMD has units of GeV.
We use $\beta=1$ and $R=R_{\max}=11.64$, the maximum angular separation in general-purpose LHC detector geometries~\cite{ATLAS:2023dns,CMS:2023gfb}.

\ifanonymousartifacts
\else
Our $Z$+jets and $t\bar t$ samples~\cite{doherty2026,doherty2026zenodo} are produced at hard-scatter (HS), parton-shower (PS), and hadronization (HAD) stages, corresponding respectively to the short-distance process, subsequent radiation due to the strong interaction, and stable final-state hadrons\footnote{A comprehensive reference on Monte Carlo event generation for collider physics is provided in Ref.~\cite{Campbell:2022qmc}.}.
\fi
The median multiplicities $(\mathrm{HS},\mathrm{PS},\mathrm{HAD})$ are $(4,29,110)$ for $Z$+jets and $(6,51,201)$ for $t\bar t$.
At HAD level, their 95th percentiles are $181$ and $301$, and their maxima are $372$ and $552$.
We train on $Z$+jets PS pairs and use the other stages and $t\bar t$ samples to test the performance of our algorithm across a wider range of particle multiplicities.
Events, rather than constructed pairs, are partitioned among training, validation, and test sets so no event is shared across splits.

\ifanonymousartifacts
\else
The processed training, validation, and test pairs are archived on the CERN Zenodo platform~\cite{leblanc_2026_22150295}, and the accompanying implementation is available on GitHub~\cite{leblanc_MAPFN_github}.
\fi

\subsection{Infrared and Collinear Safety}
\label{sec:irc-background}

Infrared and collinear (IRC) safety requires observables be insensitive to particles with vanishing transverse momentum and to a particle splitting into coincident branches with the same total momentum~\cite{Komiske:2020qhg}. A general additive IRC safe observable $\mathcal{O}(\mathcal{E})$ takes the following form \cite{Komiske:2019fks}:

\begin{equation}
    \mathcal{O}(\mathcal{E})=\sum_{i\in\mathcal{E}} p_{T,i}\psi(\vec{p}_i)
\end{equation}
where $p_T,\vec{p}$ are the particle transverse momentum and angular coordinates and $\psi$ is an $L$-Lipschitz function with constant $L$. The EMD places a bound on the difference between $\mathcal{O}(\mathcal{E}),\mathcal{O}(\mathcal{G})$ \cite{Komiske:2019fks}:

\begin{equation}
    \frac{1}{RL}\left|\mathcal{O}(\mathcal{E}) - \mathcal{O}(\mathcal{G})\right|\leq\mathrm{EMD}(\mathcal{E},\mathcal{G}).
\end{equation}
If two events $\mathcal{E},\mathcal{G}$ are nearby under the EMD, then any IRC safe observable of those events must also be close. Re-framed in a geometric language, IRC safety emerges naturally from the fact that the EMD is a continuous metric between events \cite{ba_shaper_2023}.

IRC-safety is important because low-momentum particles and nearly collinear splittings are common in collider-event representations due to the singularity structure of quantum chromodynamics, and can also be affected by simulation or detector resolution.
IRC safety makes the measurement function insensitive to the corresponding degenerate final states, permitting real and virtual singularities to cancel~\cite{10.1143/PTP.19.159,Kinoshita:1962ur,Lee:1964is}.
A set-valued regressor need not inherit IRC safety: adding or splitting a point can finitely change its latent representation.
The MA-PFN constraints introduced below do not enforce these limits, so we test both its limiting behavior and its response at finite perturbation size. 

\subsection{Metric-Aware Particle Flow Network Implementation}

The EMD is defined over two events so the MA-PFN uses the Multiset Deep Sets theorem \cite{Tabaghi_Wang_2023, Gui_Zhang_Zhong_Qiu_Wu_Ye_Wang_Liu_2019}
 \begin{equation}
    f(\mathcal{E},\mathcal{G})=F\left(\sum_{i=1}^{N_\mathcal{E}}\Phi(x_i), \sum_{j=1}^{N_\mathcal{G}}\Phi(x_j')\right),
    \label{eq:ma-pfn}
 \end{equation}
with $N_\mathcal{E}, N_\mathcal{G}$ being the multiplicities of events $\mathcal{E},\mathcal{G}$ respectively. A shared encoder $\Phi:\mathbb{R}^4\rightarrow\mathbb{R}^L$ embeds particles from each event into one of two sums $z_{\mathcal E}=\sum_{i=1}^{N_\mathcal{E}}\Phi(x_i)$ and $z_{\mathcal G}=\sum_{j=1}^{N_\mathcal{G}}\Phi(x_j')$. The particle coordinates shown to the $\Phi$ function are $p_T,\eta,\phi$ and a zeroed-out auxiliary fourth coordinate. 
The shared encoder has widths $4\to100\to100\to L$, ReLU activations after its first two affine layers, and a linear $L=64$ output.

We form symmetric and antisymmetric representations from the event latents,
\begin{equation}
    \Sigma=z_{\mathcal E}+z_{\mathcal G},
    \qquad
    \Delta=z_{\mathcal E}-z_{\mathcal G}.
    \label{eq:mapfn-pair-features}
\end{equation}
A joint MLP $F:\mathbb R^{2L}\to\mathbb R$ maps $\Sigma,\Delta$ to a scalar.
Here comma separation denotes concatenation, and $F$ has widths $128\to100\to100\to100\to1$, ReLU hidden activations, and a linear output.
We explicitly symmetrize its scalar output over the two event orientations,
\begin{equation}
    a_{\mathrm{sym}}(\Sigma,\Delta)
    =\frac{1}{2}\left[
        F(\Sigma,\Delta)+F(\Sigma,-\Delta)
    \right].
    \label{eq:mapfn-symmetric-head}
\end{equation}
The predicted distance is
\begin{equation}
    \widehat d(\mathcal E,\mathcal G)
    =\frac{\left\|\Delta\right\|_1}{L}\,
    \mathrm{softplus}\left(a_{\mathrm{sym}}(\Sigma,\Delta)\right).
    \label{eq:mapfn-output}
\end{equation}
It is useful to write the two factors in Eq.~\ref{eq:mapfn-output} as
$\rho(\mathcal E,\mathcal G)=\|z_{\mathcal E}-z_{\mathcal G}\|_1/L$ and
$g(\mathcal E,\mathcal G)=\mathrm{softplus}(a_{\mathrm{sym}}(\Sigma,\Delta))$.
The first factor, $\rho$, is an exact pseudometric on events for any encoder $\Phi$: it is non-negative and symmetric, vanishes when $z_{\mathcal E}=z_{\mathcal G}$, and identically satisfies the triangle inequality.
It is not guaranteed to be a metric only because the encoder need not be injective, so distinct events may share a latent representation (Section~\ref{sec:metric-properties}).
The second factor, $g>0$, is a learned, pair-dependent gauge that reweights this latent pseudometric.
Exchanging events leaves $\Sigma$ unchanged and maps $\Delta\to-\Delta$; by Eq.~\ref{eq:mapfn-symmetric-head}, $g$ is therefore invariant under $\mathcal E\leftrightarrow\mathcal G$.
The MA-PFN is thus a positive symmetric gauge times an exact latent pseudometric; any departure of $\widehat d$ from an exact pseudometric is carried by variation of the scalar $g$.
Although $\rho$ satisfies the triangle inequality, multiplication by the pair-dependent $g$ does not generally preserve it.
The construction also does not enforce the infrared and collinear limits, these properties are evaluated empirically in Section~\ref{sec:irc-safety}.
Additional details relating to the Multiset Deep Sets theorem may be found in Appendix \ref{app-universality}.

The baseline PFN~\cite{Komiske:2018cqr} uses the standard Deep Sets decomposition~\cite{zaheer_deep_2018}
 \begin{equation}
    f(\mathcal J)=F\left(\sum_{i=1}^{N_\mathcal{J}}\Phi(x_i)\right),
    \label{eq:pfn}
 \end{equation}
where $\mathcal{J}=\mathcal{E}\cup\mathcal{G}$ includes all particles from both events and $N_\mathcal{J}$ is the multiplicity. Constructing $\mathcal{J}$ is necessary because the standard Deep Sets theorem is a statement about functions of a single set. 
For an event pair, we decorate particles with a fourth feature $t\in\{+1,-1\}$ that identifies their parent event and apply a single encoder to $\mathcal{J}$.
The encoder $\Phi:\mathbb{R}^4\rightarrow\mathbb{R}^L$ constructs an $L$ dimensional latent representation for the event pair and the decoder $F:\mathbb{R}^L\rightarrow\mathbb{R}$ learns the mapping from latent space to output.
The hidden widths and activations of the encoder/decoder are the same as for the MA-PFN. 

Implemented in \textsc{PyTorch}~\cite{NEURIPS2019_9015}, the MA-PFN has $50{,}265$ trainable parameters and is trained on $6{,}395{,}676$ exact EMD labels, with $798{,}216$ validation pairs from the event-disjoint $Z$+jets PS split.
The corresponding `baseline' PFN has $43{,}865$ trainable parameters.
Using random seed $23{,}411$, it is optimized with AdamW~\cite{LoshchilovH19}, learning rate $10^{-4}$, zero weight decay, and global batch size $B=1024$.
The training objective balances relative and absolute accuracy,
\begin{equation}
    \mathcal L
    =\frac{1}{B}\sum_{b=1}^{B}
      \frac{|\widehat d_b-d_b|}{|d_b|+\epsilon}
      +0.25\,\frac{1}{90~\mathrm{GeV}}
      \frac{1}{B}\sum_{b=1}^{B}|\widehat d_b-d_b|.
    \label{eq:training-objective}
\end{equation}
where $\epsilon=10^{-8}$~GeV is a stabilizing factor to prevent division by 0, and $d_b$ \& $\widehat d_b$ are respectively the exact and predicted EMDs for pair $b$.
The relative weight of the MAE term was chosen before test-set evaluation in a single-seed validation ablation using the absolute-difference joint-head precursor, with the data split, initialization, and optimizer settings fixed.
We compared pure MAPE, pure MAE, and hybrid objectives with MAE weights $\lambda\in\{0.25,0.5,1\}$, where $\lambda$ replaces the coefficient $0.25$ in Eq.~\ref{eq:training-objective}.
%
%
For each run, we evaluated its best-objective checkpoint and computed an equal-weight mean rank over validation MAPE, global MAE, MAE below $35~\mathrm{GeV}$ and at or above $500~\mathrm{GeV}$, and the absolute mean residual in those two regions.
The $\lambda=0.25$ objective had the lowest balanced rank and was fixed for both the final exchange-symmetrized MA-PFN and the baseline PFN control.
%

Training stops after 50 epochs without improvement in the validation objective or at 700 epochs.
The selected run stopped after epoch 698, and we use the epoch-648 checkpoint that minimizes the validation objective in Eq.~\ref{eq:training-objective}.


\section{Results}

\subsection{Predictive Accuracy}

We evaluate predictive accuracy on $798{,}216$ held-out $Z$+jets PS event pairs, using the exact POT calculation as the reference and using FP32 predictions for both networks.
In this sample, the pairwise EMD has a mean of $114.9$ GeV and a median of $88.84$ GeV; its central $90\%$ spans $27.0$–$292.4$ GeV, with a full range of $3.79$–$810.4$ GeV.
Across the full sample, the MA-PFN achieves an MAE of $0.679$~GeV, an RMSE of $0.943$~GeV, and a mean absolute percentage error (MAPE) of $0.890\%$.
The corresponding baseline PFN values are $0.843$~GeV, $1.176$~GeV, and $1.124\%$, respectively.
Thus, the metric-aware constraints do not reduce regression accuracy; the MA-PFN instead lowers each of these errors by approximately $20\%$ relative to the matched baseline PFN.

\begin{figure}[htbp]
    \centering
    \begin{subfigure}{0.49\textwidth}
        \centering
        \includegraphics[width=\linewidth]{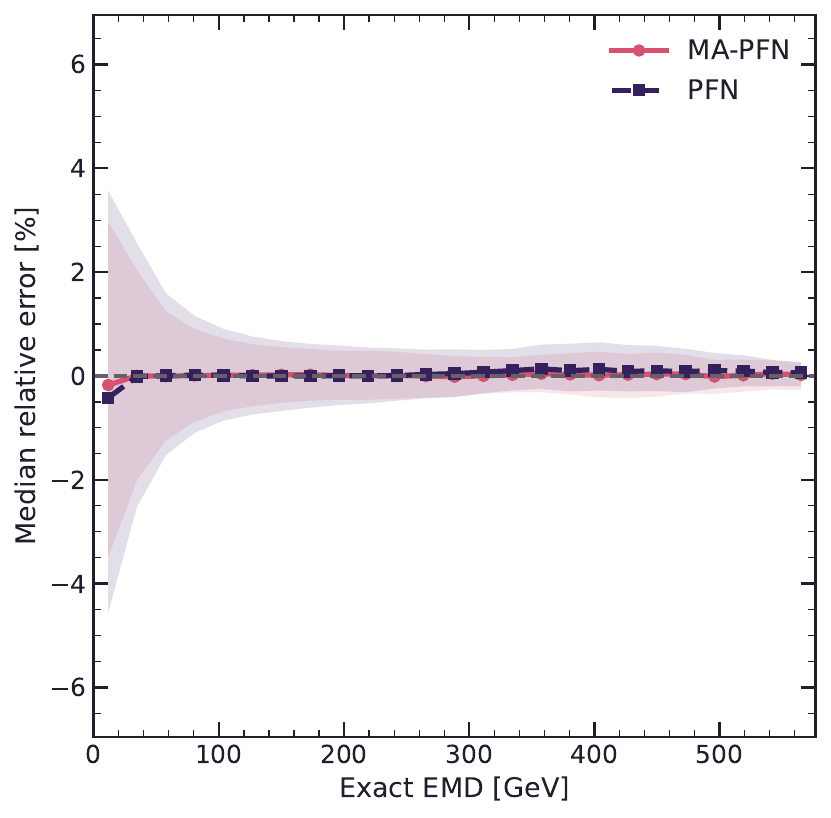}
        \caption{Relative residual}
        \label{fig:predictive-relative-error}
    \end{subfigure}
    \begin{subfigure}{0.49\textwidth}
        \centering
        \includegraphics[width=\linewidth]{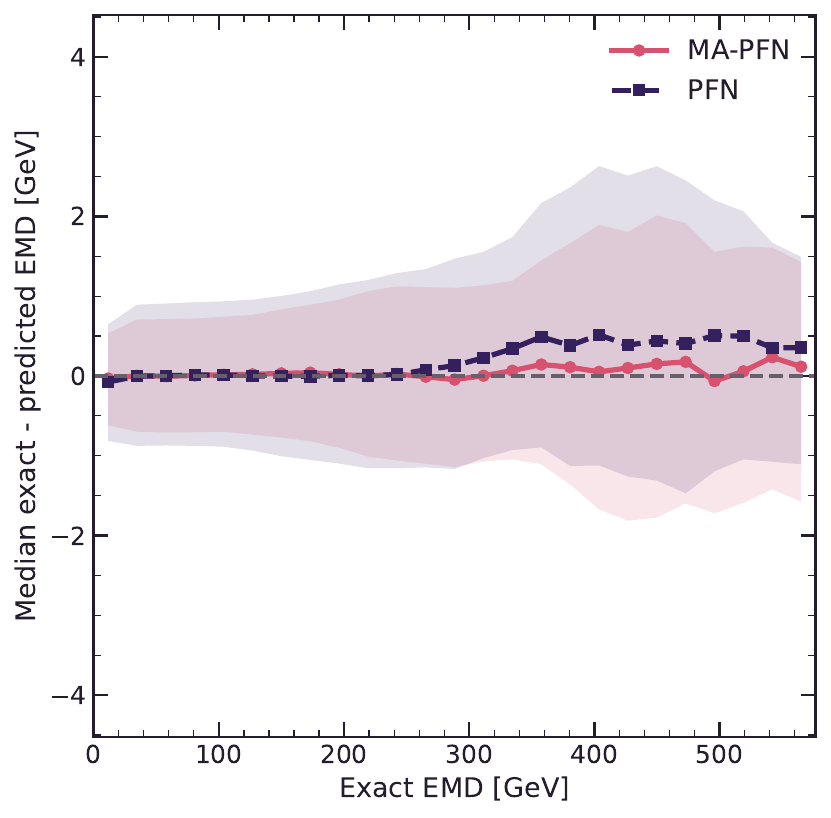}
        \caption{Residual}
        \label{fig:predictive-residual}
    \end{subfigure}
    \caption{Predictive residuals as functions of the exact EMD.
            Curves show the binned median, and shaded bands indicate the central $68\%$ interval.
            }
    \label{fig:predictive-response}
\end{figure}

We define the signed residual as $r=d-\widehat d$ and the signed relative residual as $100r/d$.
Figures~\ref{fig:predictive-response}\subref{fig:predictive-relative-error} and~\ref{fig:predictive-response}\subref{fig:predictive-residual} show their binned medians and central $68\%$ intervals as functions of the exact EMD.
The MA-PFN median remains close to zero and is generally smaller than that of the baseline PFN, particularly at larger EMD.
The absolute residual spread grows with EMD, whereas the relative spread is largest at small EMD because a comparable absolute error is divided by a smaller distance.
Downstream applications where small EMD performance is important should be mindful of the larger fractional uncertainty observed from both neural surrogate networks in this region.

\subsection{Learned Metric Properties}\label{sec:metric-properties}

The MA-PFN enforces non-negativity, zero self-distance, and exchange symmetry by construction (up to finite-precision effects), whereas the triangle inequality is not imposed explicitly.
We evaluate these properties on an independent sample of $93{,}895$ held-out events not used for training, validation, or testing.
The exchange-symmetry and triangle-inequality studies use $10^6$ randomly sampled pairs and triplets, respectively. 
All comparisons use a tolerance of $10^{-3}$~GeV, which is far smaller than the typical minimum energy requirement of a reconstructed particle in LHC data analysis (typical thresholds range from $0.1--0.5$~GeV for reconstructed charged-particle tracks and calorimeter cell clusters).
No negative distances are observed for either the MA-PFN or baseline PFN.

\begin{figure}[htbp]
    \centering
    \begin{subfigure}{0.49\textwidth}
        \centering
        \includegraphics[width=\linewidth]{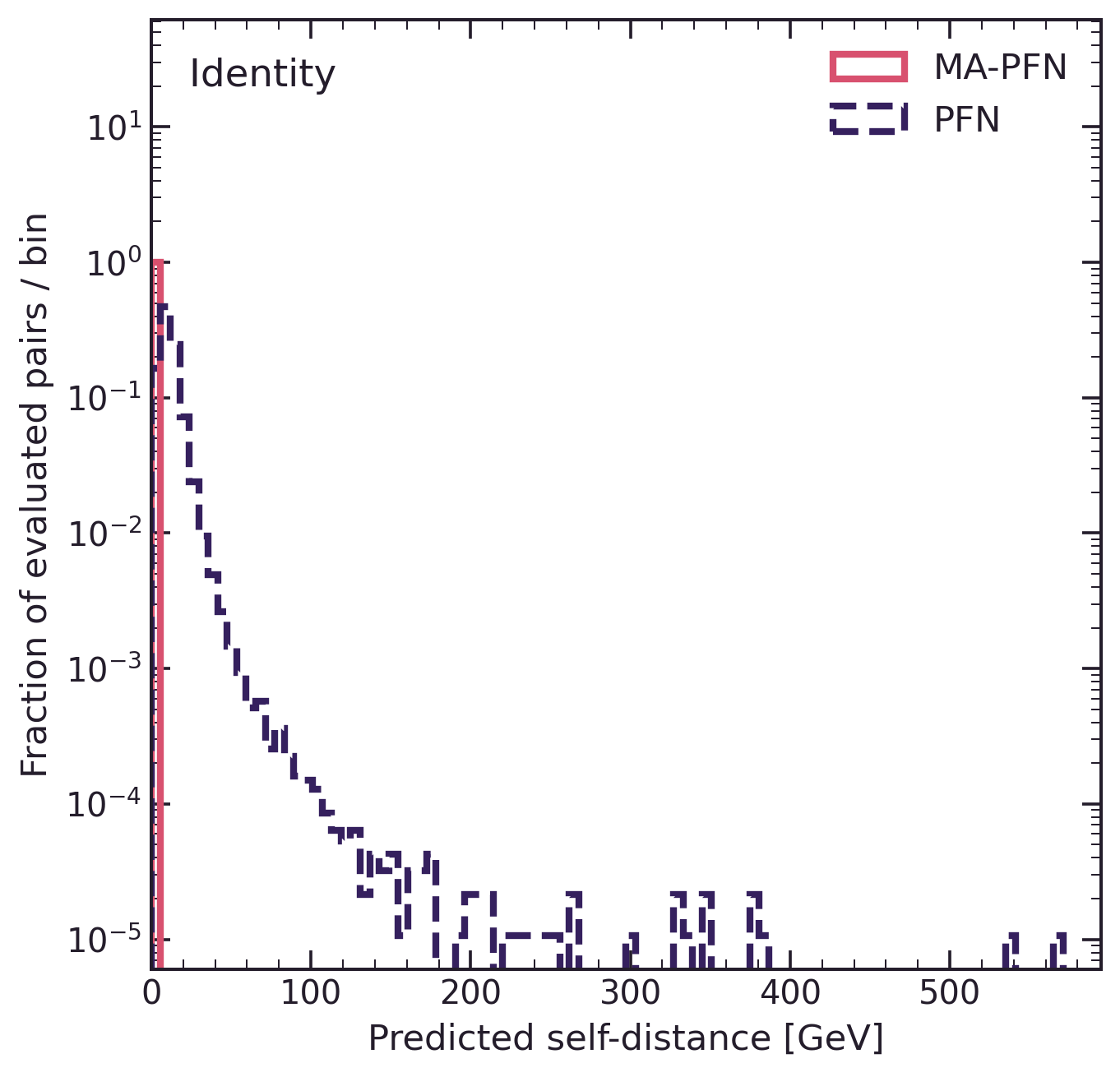}
        \caption{Identity}
        \label{fig:metric-identity}
    \end{subfigure}
    \begin{subfigure}{0.49\textwidth}
        \centering
        \includegraphics[width=\linewidth]{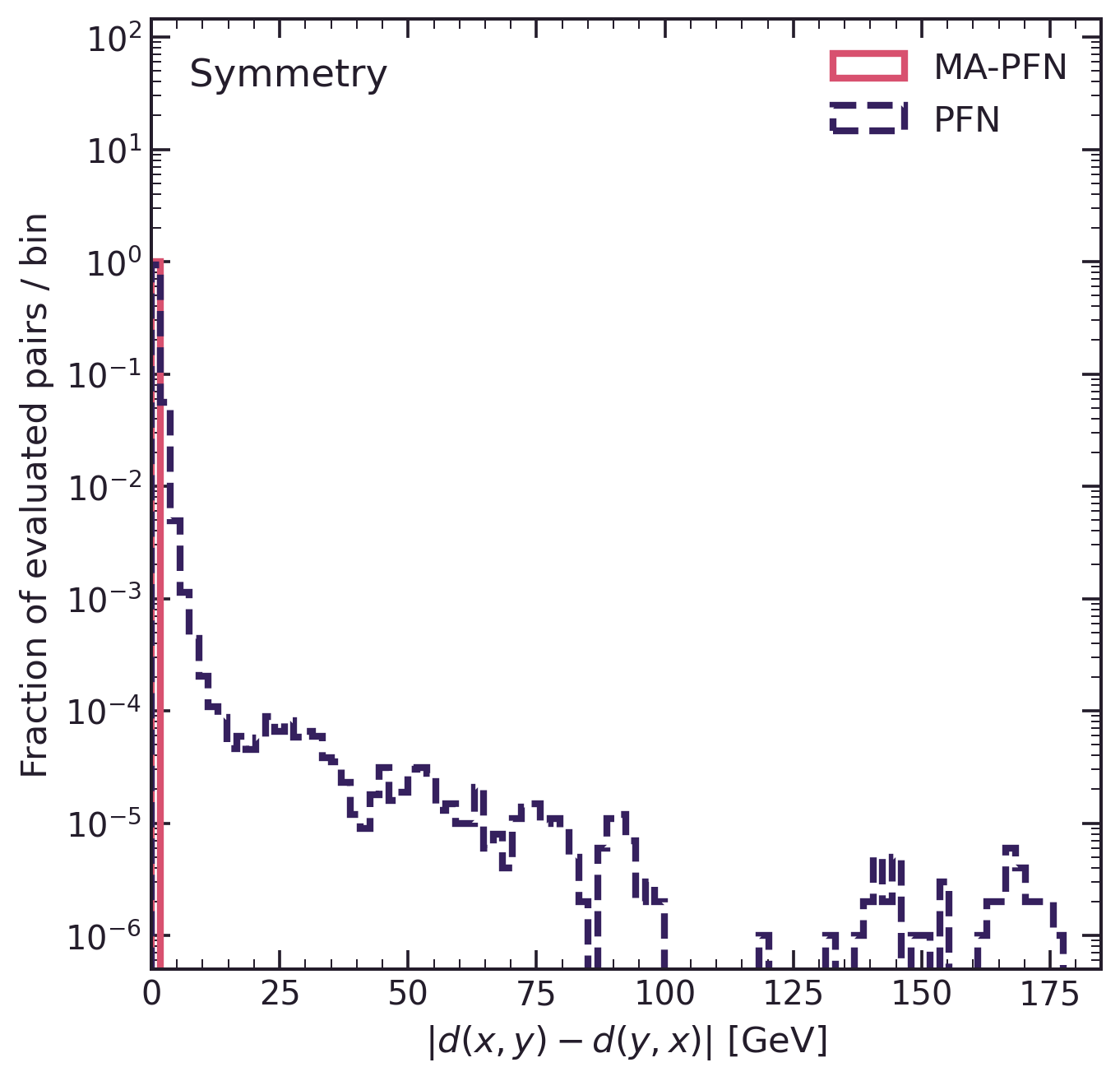}
        \caption{Exchange symmetry}
        \label{fig:metric-symmetry}
    \end{subfigure}
    \begin{subfigure}{0.49\textwidth}
        \centering
        \includegraphics[width=\linewidth]{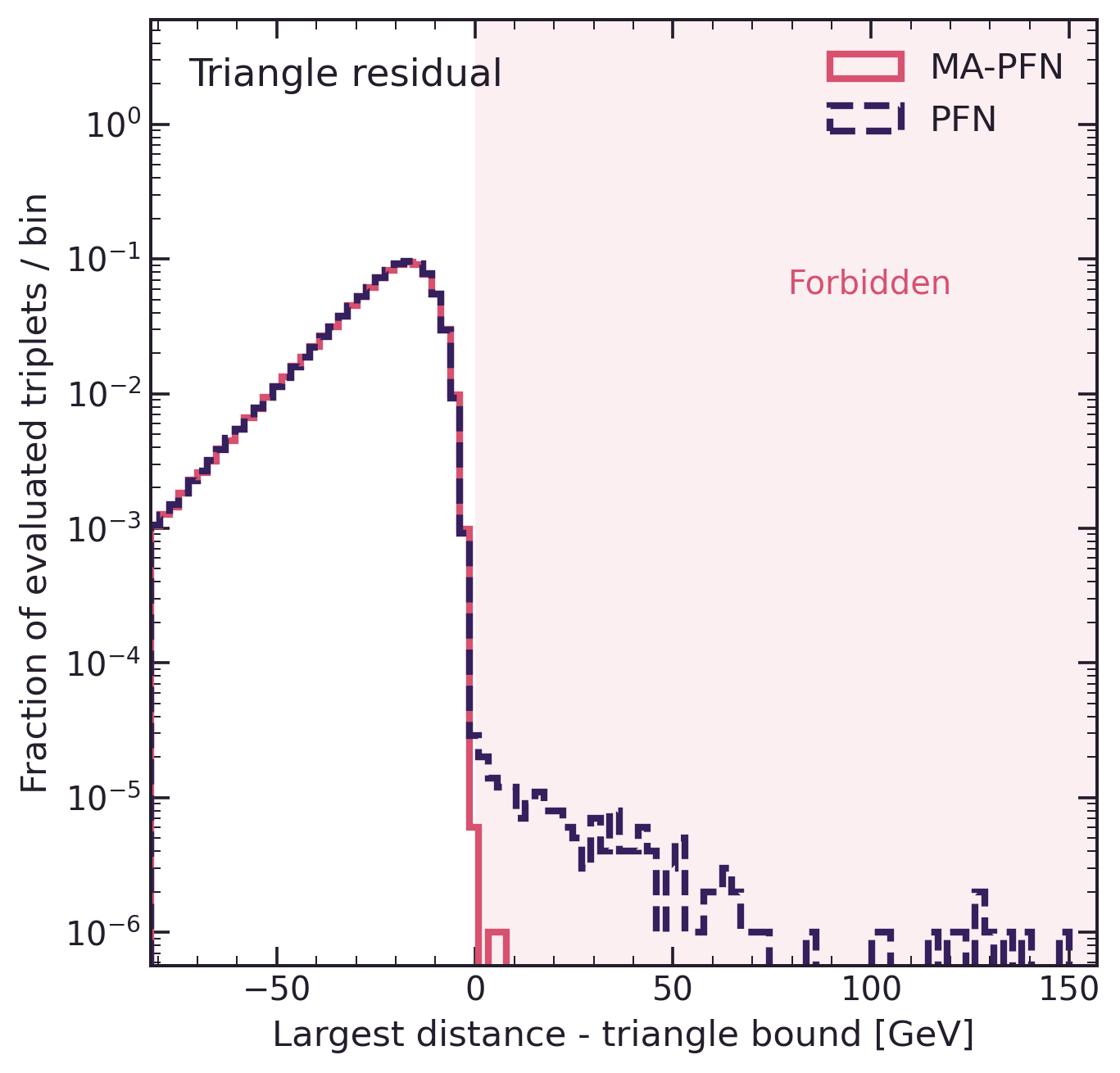}
        \caption{Triangle inequality}
        \label{fig:metric-triangle}
    \end{subfigure}
    \caption{Metric-property benchmarks for the MA-PFN and matched baseline PFN.}\label{metric-benchmarks}
\end{figure}

Figure~\ref{metric-benchmarks}\subref{fig:metric-identity} shows the predicted self-distance.
The MA-PFN returns exactly zero for every event, while none of the baseline PFN self-distances lies within  numerical tolerance of zero; the tail of the baseline distribution extends to $566$~GeV.
Although the architecture guarantees \(\hat d(x,x)=0\), it does not enforce the converse.
Among \(10^6\) sampled pairs of distinct held-out events, none had \(\hat d(x,y)\leq10^{-3}\) GeV; the minimum predicted distance between distinct pairs for both the MA-PFN and baseline PFN was \(4.21\) GeV.
Thus, no empirical violations of identity of indiscernibles were observed at the stated tolerance. 
%
%
Similarly, all MA-PFN predictions in Figure~\ref{metric-benchmarks}\subref{fig:metric-symmetry} satisfy exchange symmetry within the numerical tolerance, with a maximum discrepancy of $6.1\times10^{-4}$~GeV.
For the baseline PFN, only $0.103\%$ of the evaluated pairs satisfy the same tolerance, and the maximum discrepancy is $176$~GeV.

To test the triangle inequality, we define
\[
r_\triangle = d_{\max}-(d_1+d_2),
\]
where $d_{\max}$ is the largest of the three predicted distances and $d_1$ and $d_2$ are the other two.
Values of $r_\triangle$ greater than the numerical tolerance constitute violations.
As shown in Figure~\ref{metric-benchmarks}\subref{fig:metric-triangle}, only 2 of the $10^6$ MA-PFN triplets violate the inequality, compared with 199 for the baseline PFN, corresponding to an approximately hundredfold reduction in the observed violation rate.
The largest positive residual is also reduced from $149.5$~GeV for the baseline PFN to $5.8$~GeV for the MA-PFN.
The two remaining MA-PFN violations emphasize that triangle consistency is learned empirically rather than guaranteed by the architecture.

\subsection{Learned Infrared and Collinear Safety}
\label{sec:irc-safety}

To probe the infrared and collinear safety of this algorithm as described in Section~\ref{sec:irc-background}, we sample $10^4$ held-out pairs $(\mathcal E,\mathcal G)$ with distinct events and perturb only $\mathcal E$.
For each perturbation scale $\xi$, we compare the change in predicted distance with the exact EMD response,
\begin{equation}
    r_D(\xi)
    =100\,
    \frac{\left|D(\mathcal E'(\xi),\mathcal G)-D(\mathcal E,\mathcal G)\right|}
    {\mathrm{EMD}(\mathcal E,\mathcal G)},
    \qquad D\in\{\mathrm{EMD},\widehat d\}.
    \label{eq:paired-irc-response}
\end{equation}
Using distinct reference events avoids normalization by a vanishing self-distance and permits direct comparison with the exact EMD response.
The infrared path adds a particle at $(\eta,\phi)=(0,0)$, while the collinear path replaces the leading particle with two equal-$p_T$ branches separated in periodic $\phi$ while preserving their $p_T$-weighted centroid on the unwrapped coordinate.
Both scans use 60 logarithmically spaced points between $10^{-3}$ and $10$.

\begin{figure}[htbp]
    \centering
    \begin{subfigure}{0.49\textwidth}
        \centering
        \includegraphics[width=\linewidth]{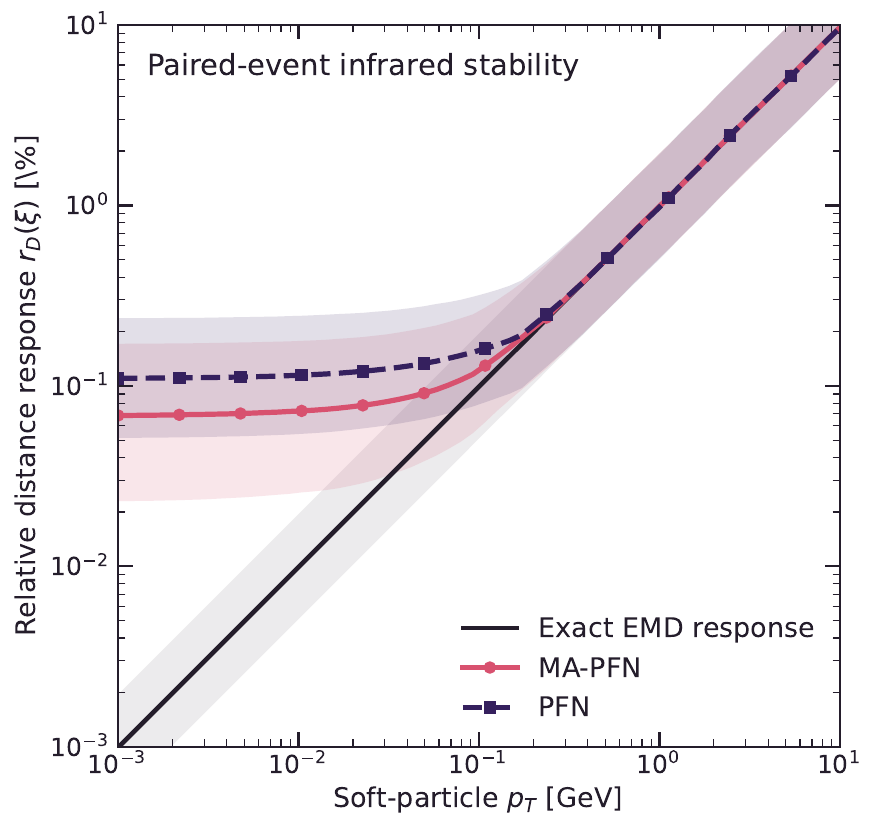}
        \caption{Soft response}
        \label{fig:irc-soft-response}
    \end{subfigure}
    \begin{subfigure}{0.49\textwidth}
        \centering
        \includegraphics[width=\linewidth]{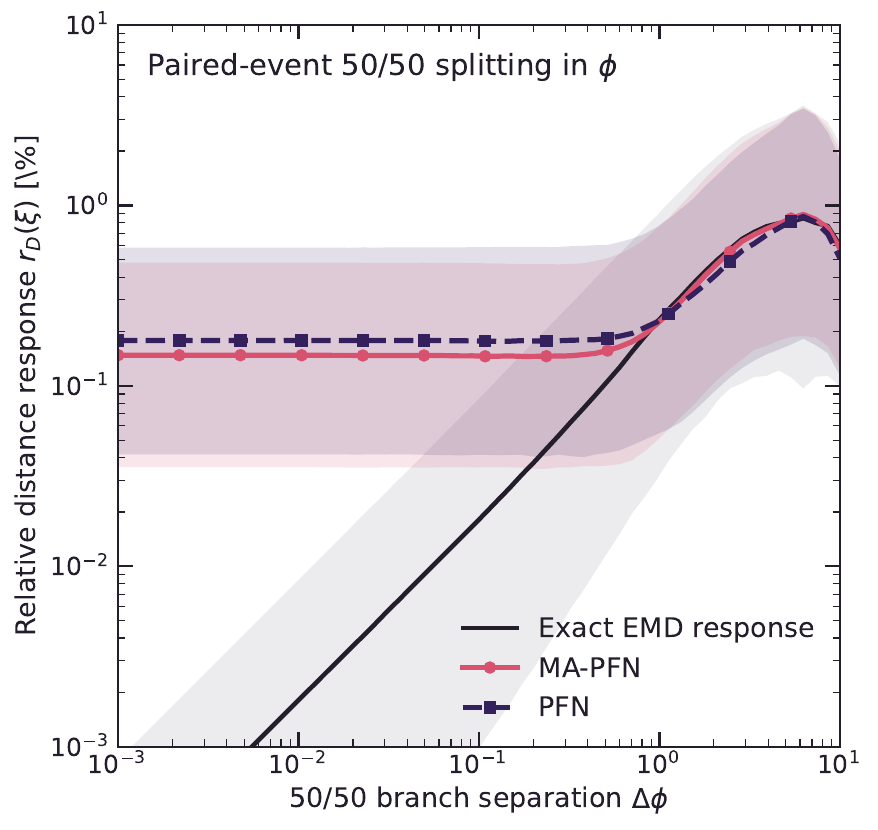}
        \caption{Collinear response}
        \label{fig:irc-collinear-response}
    \end{subfigure}
    \caption{Relative distance response for $10^4$ distinct held-out event pairs under soft emission and a 50/50 collinear splitting in $\phi$.
    Curves show medians and bands the 16th--84th percentiles.
    %
    %
    %
    }
    \label{fig:irc-response}
\end{figure}

Figure~\ref{fig:irc-response} shows event-level medians and central $68\%$ intervals for Eq.~\ref{eq:paired-irc-response}.
In the infrared test, the exact response decreases approximately linearly with the added particle's $p_T$, while both learned models approach a nonzero floor.
The MA-PFN follows the exact response to smaller values before plateauing, implying that it has learned a more expressive representation than the baseline PFN.
%
%

The $50\%/50\%$ collinear splitting exhibits a similar small-angle floor.
The large-angle peak near $\Delta\phi=2\pi$ is a consequence of azimuthal periodicity rather than collinear behavior: at this nominal separation, the two daughters wrap to the same point opposite the original particle, producing their maximum displacement.
Only the $\Delta\phi\to0$ region should therefore be interpreted as a test of collinear safety.
In this case, the MA-PFN also follows the exact response to slightly lower values than the baseline PFN before plateauing.
This behaviour was consistently observed also in uneven splittings with $10\%/90\%$ and $25\%/75\%$ (not illustrated).

\subsection{Timing Performance}

The main timing workload takes two disjoint sets of $1{,}000$ events and evaluates all $10^6$ cross-pairs.
We compare the exact POT 0.9.7.post1~\cite{flamary2026pot,flamary2021pot}, Wasserstein C++ 1.1.0~\cite{10.1145/2024156.2024192,Komiske:2019fks,Komiske:2020qhg}, and EnergyFlow.jl 0.1.0 \texttt{ns64} and \texttt{ot64} backends~\cite{leblanc_2026_22049386} with the approximate methods MA-PFN, a SHAPER-style GeomLoss 0.2.6 Sinkhorn divergence~\cite{feydy2019interpolating,ba_shaper_2023}, and a custom ten-projection planar sliced Wasserstein distance~\cite{bonneel_sliced_2015}.
We do not include iterative differentiable solvers such as NEEMo~\cite{Kitouni:2022qyr}, which optimize a pair-dependent dual potential and target geometric fitting rather than amortized evaluation of dense event--event distance matrices.
Canonical event arrays are loaded and selected before measurement.
For every method, wall timing begins immediately before method-specific pair preparation and ends only after the ordered array of $10^6$ predictions is resident in host memory.
This means that timing of disk input, imports, solver and worker construction, and JIT warm-up are excluded; such effects are expected to be negligible at the scale of a realistic full-scale LHC data analysis or MC production.

The performance of MA-PFN is measured with a cold input cache.
Within the timed interval, both event sets and the $10^6$ pair indices are transferred to the GPU, each of the $2{,}000$ unique events is encoded once, and the resulting $64$-dimensional latents and pair indices remain GPU-resident while the joint head evaluates all pairs in batches.
The interval also includes latent indexing, CUDA synchronization, and transfer of the predictions back to host memory.
This procedure avoids the need to re-encode each event for all $1{,}000$ pairwise calculations in which it participates, without excluding input preparation or encoding from the reported time.
As summarized in Figure~\ref{fig:pred_and_timing}\subref{fig:million-pair-multiplicity}, mixed-precision MA-PFN takes $0.038$--$0.057$~s on one NVIDIA A40, corresponding to $17$--$26$ million pairs per second.
The MA-PFN wall time changes only modestly across the sampled multiplicity range in Figure~\ref{fig:pred_and_timing}\subref{fig:million-pair-multiplicity} because each event is encoded once, whereas the exact-solver curves rise much more steeply with event multiplicity.
Relative to the fastest exact implementation on $48$ physical CPU cores for each sample, the measured throughput advantage grows from $1.5\times$ for median multiplicities of $4$ particles, to $1.8\times10^3$ for median multiplicities of $202$ particles.
As these numbers are not normalized to hardware~\cite{Michelotto:2010zz}, some care must be taken with their interpretation.
In particular, configurations with 48 CPUs for single analysis or production workflows at the LHC are currently uncommon, while those with either a dedicated GPU or accelerators available as a service~\cite{Duarte:2019fta,Krupa:2020bwg} are becoming more commonplace.
Complete implementation, batching, and repetition details are given in Appendix~\ref{app:timing-study}.

\subsubsection{Comparison of approximate methods}

\begin{figure}[htbp]
    \centering
    \begin{subfigure}{0.49\textwidth}
        \centering
        \includegraphics[width=\linewidth]{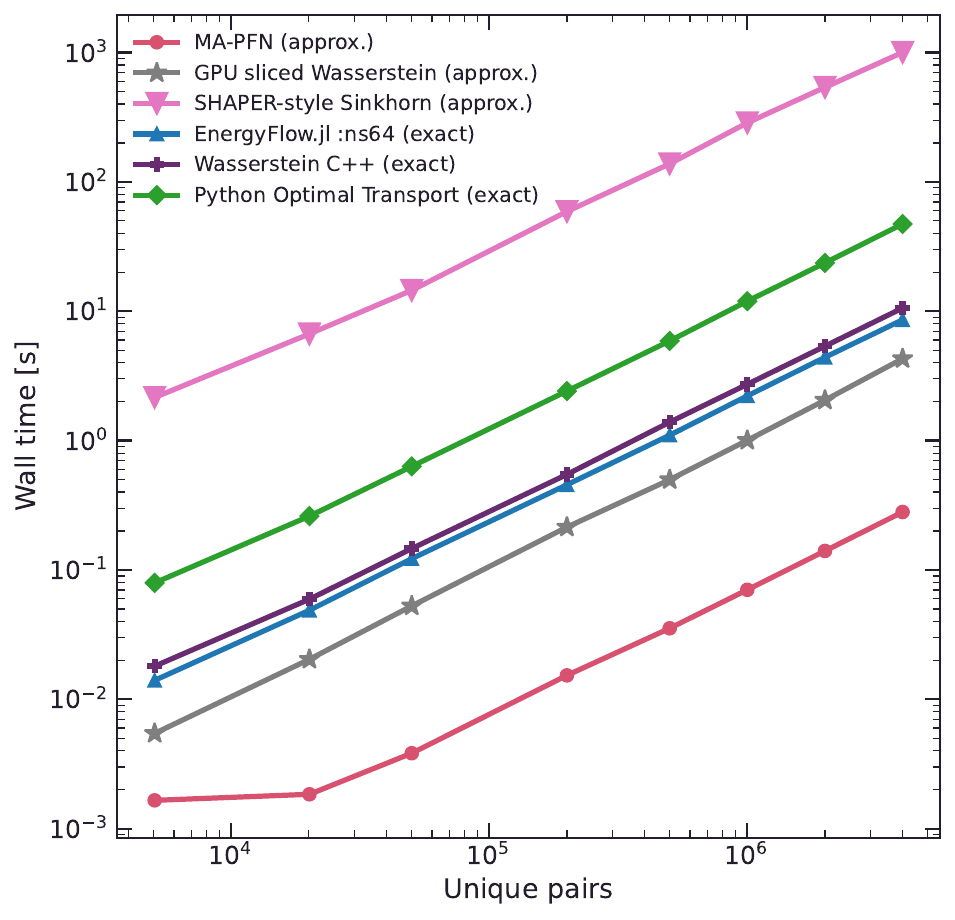}
        \caption{Pair-count scaling}
        \label{fig:pair-count-scaling}
    \end{subfigure}
    \begin{subfigure}{0.49\textwidth}
        \centering
        \includegraphics[width=\linewidth]{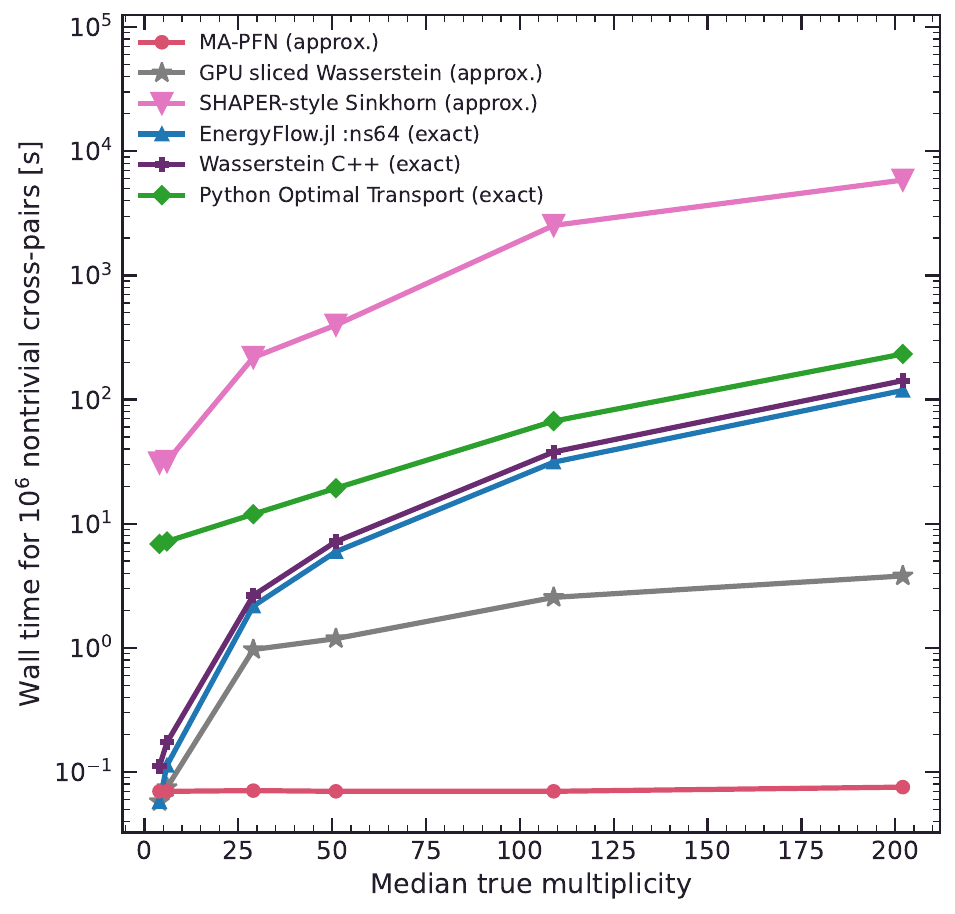}
        \caption{Multiplicity scaling at $10^6$ pairs}
        \label{fig:million-pair-multiplicity}
    \end{subfigure}\\
  \caption{
    Wall-time scaling of the MA-PFN and alternative EMD methods.
    The pair-count study uses the $Z$+jets PS sample; the multiplicity study evaluates $10^6$ cross-pairs in each of the six event samples.
    MA-PFN timing includes event transfer, unique-event
    encoding, pairwise evaluation, synchronization, and transfer of the predictions back to host memory.
    }
    \label{fig:pred_and_timing}
\end{figure}

\begin{figure}[htbp]
        \begin{subfigure}[t]{0.49\textwidth}
        \centering
            \includegraphics[width=\linewidth]
            {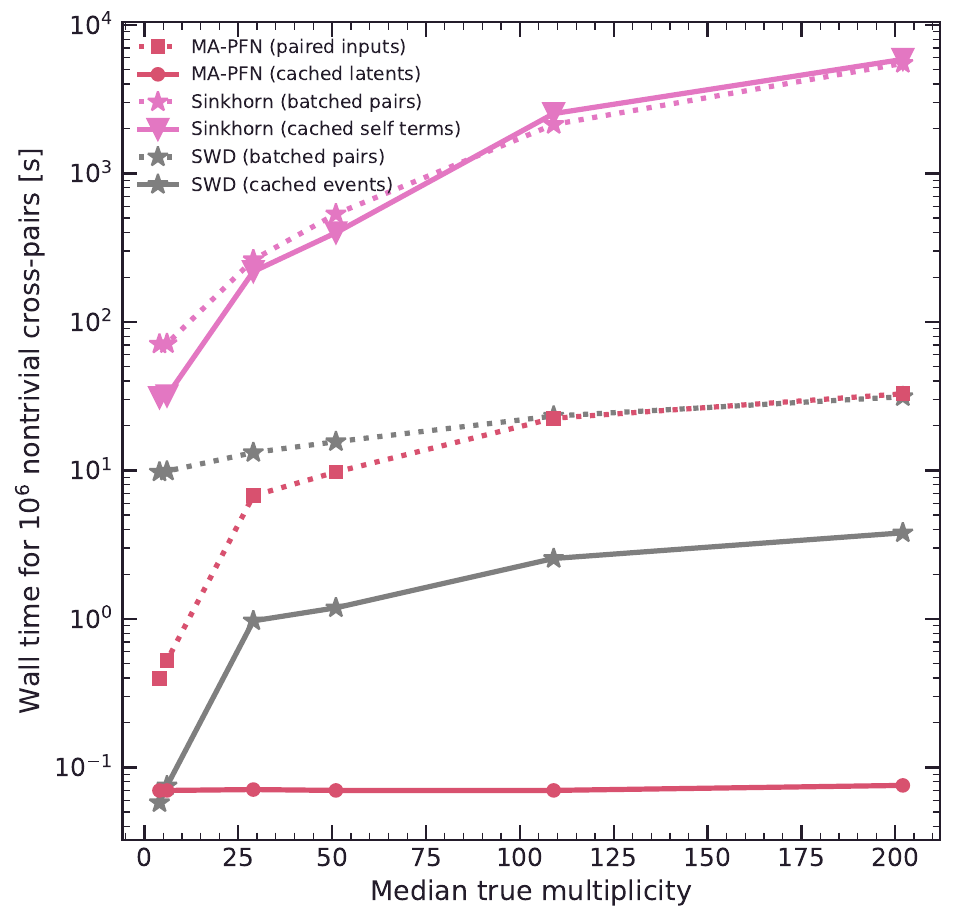}
            \caption{Implementation ablation}
            \label{fig:implementation-ablation}
        \end{subfigure}
        \begin{subfigure}[t]{0.49\textwidth}
        \centering
            \includegraphics[width=\linewidth]
            {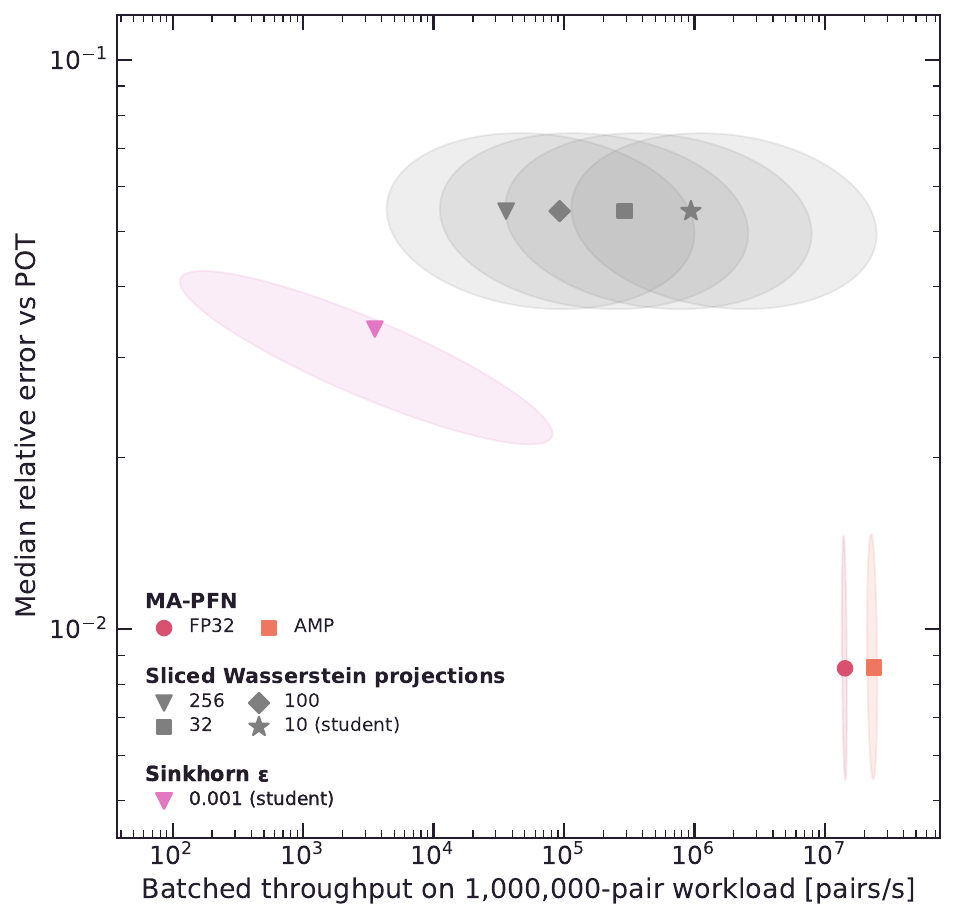}
            \caption{Accuracy--throughput frontier}
            \label{fig:approximate-frontier}
        \end{subfigure}
      \caption{
    Implementation and accuracy--throughput comparisons for the approximate methods.
    (\subref{fig:pair-count-scaling}) The implementation ablation reports wall time for $10^6$ cross-pairs as a function of median particle multiplicity.
    Cached implementations reuse per-event state; those curves are compared with matched setups that rebuild the per-event state for every pair.
    (\subref{fig:million-pair-multiplicity}) The accuracy--throughput frontier compares median absolute relative error with resident-cache throughput on $10^6$-pair workloads.
    Large markers give the median across the six event samples, small markers show the individual samples, and shaded ellipses summarize their dispersion.
    Lower and farther right is preferred.
    }
    \label{fig:ablation_and_pareto}
\end{figure}

Figure~\ref{fig:ablation_and_pareto} isolates the implementation choices responsible for the performance of the approximate methods and compares their accuracy--throughput tradeoffs on a workflow of $10^6$ pairwise EMD calculations.
The MA-PFN factorization is well suited to a dense pairwise workload: each of the $2{,}000$ unique events is encoded once and stored in GPU memory, rather than being computed once for every pair in which it appears.
Thus, the multiplicity-dependent constituent processing is paid only during the $\mathcal{O}(Nn)$ event-encoding stage, where $N$ is the number of events per set and $n$ is particle multiplicity. The $10^6$ comparisons are then reduced to an $\mathcal{O}(N^2)$ pair stage consisting of fixed-width head evaluations on cached 64-dimensional latents, independent of particle multiplicity.
This amortization explains the nearly flat MA-PFN curve in Figures~\ref{fig:pred_and_timing}\subref{fig:million-pair-multiplicity}~and~\ref{fig:ablation_and_pareto}\subref{fig:implementation-ablation}.

The same reuse principle can improve the performance of other approximate methods, as illustrated in Fig.~\ref{fig:ablation_and_pareto}\subref{fig:implementation-ablation}.
For the ten-projection SWD implementation, caching each event's projections and sorted supports gives the largest improvement at low multiplicity.
The remaining weighted-CDF merge is still performed for each projection and pair, so the wall time continues to grow with multiplicity.
For the Sinkhorn divergence, caching the two debiasing self terms avoids repeated self-solves, but the cross-event transport problem remains.
This gives an appreciable speedup at low and intermediate multiplicity, although the final performance does not match that of the MA-PFN with cached latents.

The accuracy--throughput Pareto frontier for the studied approximate methods is provided in Figure~\ref{fig:ablation_and_pareto}\subref{fig:approximate-frontier}.
The results of six trials per method and configuration are shown as a shaded ellipse on the plane, to provide a visualization of the trial-to-trial stability of these results.
MA-PFN was benchmarked in both FP32 and Automatic Mixed Precision (AMP) modes, which are found to have lower median absolute relative error and greater aggregate throughput than the tested SWD and Sinkhorn configurations.
Increasing the number of SWD projections produces no material improvement in the median error, but incurs the expected throughput cost.
Thus both MA-PFN precision modes lie on the tested Pareto frontier: AMP provides approximately $1.7\times$ greater median throughput, while FP32 has marginally lower median error.
The Sinkhorn configuration shown, with $\epsilon=10^{-3}$, lies between MA-PFN and SWD in accuracy, with a median absolute relative error of $3.4\%$ and reaching a median throughput of only $3.5\times10^3$ pairs/s.
It is therefore Pareto-dominated by both MA-PFN precision modes, which reduce the median error to approximately $0.85\%$ while processing at least $1.4\times10^7$ pairs/s.

\FloatBarrier
\section{Conclusion}

In this work, we introduce the Metric-Aware Particle Flow Network (MA-PFN), a permutation-invariant neural surrogate for the Energy Mover's Distance that enforces non-negativity, exchange symmetry, and zero self-distance by construction.
The model achieves median relative errors of order $1\%$ across both collision processes and all three event-generation stages considered.
On the tested hardware, it provides higher measured throughput than exact CPU optimal-transport implementations in large pairwise workloads, with an advantage that increases rapidly with event multiplicity.
Its advantage over exact solvers grows rapidly with event multiplicity, making it well suited to large pairwise studies of high-multiplicity collider physics events.

The architectural constraints also improve geometric properties that are not explicitly guaranteed.
On $10^6$ held-out event triplets, the MA-PFN reduces the observed triangle-inequality violation rate by $\sim100\times$ relative to a baseline PFN, and reduces the largest violation from $149.5$ to $5.8$~GeV.
In paired-event infrared and 50/50 collinear studies, it also has a lower median signed-response error than the PFN at every sampled perturbation scale and a smaller representation-limited response floor.
Nevertheless, both models approach nonzero plateaus in the infrared and collinear limits rather than the exact EMD response.
These results indicate that encoding a subset of the target geometry can improve the broader geometric fidelity of a learned distance, beyond the properties enforced directly.

The MA-PFN is nevertheless an approximate surrogate rather than an exact replacement for the EMD.
Separation of distinct events, the triangle inequality, and IRC safety are not guaranteed, and relative errors increase in the small-distance regime.
Future work should investigate architectures that enforce the complete metric structure, targeted training on difficult triplets and small distances, and the effect of residual geometric violations on downstream applications such as nearest-neighbor searches, clustering, and anomaly detection.


\ifanonymousartifacts

\else

\section{Acknowledgments}
We thank Rikab Gambhir for quick, insightful feedback on the draft manuscript.
%
This material is based on work supported by the U.S. Department of Energy, Office of Science, Office of High Energy Physics under Award Number DE-SC0026285; it received support in its early stages in the form of a seed grant from the Brown University Data Science Institute.
This work is supported by the National Science Foundation under Cooperative Agreement PHY-2019786 (The NSF AI Institute for Artificial Intelligence and Fundamental Interactions, http://iaifi.org/).
This research was conducted using computational resources and services provided by the Center for Computation and Visualization, Brown University. 

\fi

\bibliographystyle{unsrtnat}
\bibliography{refs}


\clearpage
\appendix
\section{Timing-study protocol and additional results}
\label{app:timing-study}

\subsection{Hardware and execution settings}

Exact solvers run in Float64 on one node of Brown University's Oscar cluster containing two Intel Xeon Platinum 8268 processors.
All $48$ physical cores are used with simultaneous multithreading disabled: POT uses $48$ single-threaded worker processes, Wasserstein C++ uses $48$ OpenMP threads, and each EnergyFlow.jl backend uses $48$ Julia threads with BLAS and OpenMP restricted to one thread.
GPU methods run sequentially within each allocation on one NVIDIA A40 with $46{,}068$~MiB of memory and eight CPU cores for staging.
The GPU measurements span multiple physical A40s of the same model and therefore compare the tested deployments rather than kernels executed on one identical device.

For all methods, timing begins after the canonical event arrays have been loaded and selected, and includes method-specific preparation, distance evaluation, synchronization, and production of an ordered host-resident output array.
Disk input, imports, environment and solver construction, worker startup, and JIT warm-up are excluded.
MA-PFN is benchmarked in FP32 and automatic mixed precision (AMP); Figure~\ref{fig:pred_and_timing}\subref{fig:million-pair-multiplicity} reports FP32, while Figure~\ref{fig:ablation_and_pareto}\subref{fig:approximate-frontier} includes both precision modes.

GPU batch sizes are tuned independently for each method and sample, reflecting the different per-pair working sets of the three GPU methods.
Candidates using more than $80\%$ of device memory are rejected, and the smallest batch size attaining at least $98\%$ of the peak median throughput is selected.
MA-PFN uses $65{,}536$ pairs except for $t\bar t$ HAD in FP32 and both HAD samples in AMP, where it uses $32{,}768$.
For ten-projection SWD, the selected sizes are $65{,}536$ for $Z$+jets HS, $32{,}768$ for $t\bar t$ HS, $16{,}384$ for both PS samples, and $4{,}096$ for both HAD samples.
Sinkhorn uses $2{,}048$ pairs for the HS and PS samples, $512$ for $Z$+jets HAD, and $256$ for $t\bar t$ HAD.
These differences arise because an MA-PFN pair evaluation operates only on two fixed-width cached latents, whereas SWD retains multiplicity-dependent projected supports and Sinkhorn constructs transport tensors whose size grows approximately with the product of the two event multiplicities.
Consequently, the higher-multiplicity HAD samples require progressively smaller SWD and especially Sinkhorn batches, while the MA-PFN pair-stage memory is nearly independent of particle multiplicity.

\subsection{Repetitions and numerical exceptions}

For the main $10^6$-pair workload, CPU methods use between three and seven repetitions to assess trial variance: at least three are performed, followed by additional repetitions up to seven unless the cumulative measured time reaches $30$~s.
SWD follows the same adaptive schedule, Sinkhorn uses three repetitions, and MA-PFN uses five repetitions for each precision and sample.
For cache-aware GPU methods, the first repetition constructs and transfers the reusable per-event state, while subsequent repetitions reuse it.
The reported MA-PFN median is a resident-cache measurement in every sample; the corresponding cold-cache preparation time is recorded separately.
Figures report median wall time, and throughput is defined as the pair count divided by that wall time.

The default Wasserstein C++ solver failed for one pair in the $t\bar t$ PS million-pair workload.
The plotted value uses an otherwise identical recovery configuration with its small-cost tolerance increased by a factor of ten, which passed the exact-solver validation.
EnergyFlow.jl \texttt{ns64} returned one non-optimal result at each of the two largest pair-count workloads, containing $2{,}001$ and $2{,}829$ events; those points are omitted.


\section{Additional Details about Multiset Deep Sets Theorem}
\label{app-universality}
The MA-PFN is constructed on top of the Multiset Deep Sets theorem \cite{Tabaghi_Wang_2023, Gui_Zhang_Zhong_Qiu_Wu_Ye_Wang_Liu_2019}. The standard Deep Sets theorem \cite{zaheer_deep_2018, Komiske:2018cqr} states that if there is a bounded set $X$ with $N$ elements $x_1,x_2,...,x_N \in \mathbb{R}^d$, a bounded interval $Y\subset \mathbb{R}$, and a function $f:X\rightarrow Y$ that is invariant under permutations of the elements of $X$, then $f$ can be approximated by the following expression:

\begin{equation}
    f(X)=F\left(\sum_{i=1}^N\Phi(x_i)\right)
    \label{deepsets}
\end{equation}
and there exists a large enough integer $L$ and continuous functions $\Phi:\mathbb{R}^d\rightarrow\mathbb{R}^L$ and $F:\mathbb{R}^L\rightarrow\mathbb{R}$ such that 
\ref{deepsets} is an arbitrarily good approximation.

Because the EMD is defined across two events, this theorem alone is insufficient to describe distances between measures. Additional modifications to the input, such as our auxiliary fourth coordinate, can partially mitigate this issue. In general, extending the Deep Sets framework to include observables defined over many independent sets is a more natural approach \cite{Tabaghi_Wang_2023, Gui_Zhang_Zhong_Qiu_Wu_Ye_Wang_Liu_2019}. 
Suppose there are $\mathcal{N}$ sets $X_1,X_2,...,X_\mathcal{N}$ each with elements $\{x_1^{X_1},x_2^{X_1},...,x_{N_1}^{X_1}\}$, $\{x_1^{X_2},x_2^{X_2},...,x_{N_2}^{X_2}\},..., \{x_1^{X_\mathcal{N}},x_2^{X_\mathcal{N}},...,x_{N_\mathcal{N}}^{X_\mathcal{N}}\}$. The multiset function takes the following form: 

\begin{align}
Y &= f(X_1,X_2,...,X_\mathcal{N}) \\
  &= f(\{\underbrace{x_1^{X_1},x_2^{X_1},...,x_{N_1}^{X_1}}_{\pi_{N_1}}\}, \{\underbrace{x_1^{X_2},x_2^{X_2},...,x_{N_2}^{X_2}}_{\pi_{N_2}}\},..., \{\underbrace{x_1^{X_\mathcal{N}},x_2^{X_\mathcal{N}},...,x_{N_\mathcal{N}}^{X_\mathcal{N}}}_{\pi_{N_\mathcal{N}}}\})
\end{align}
with $\pi_N\in S_N$, and $S_N$ being the symmetric group of $N$ elements, and $f$ is invariant to permutations within any set. The Multiset Deep Sets theorem takes the following form \cite{Tabaghi_Wang_2023, Gui_Zhang_Zhong_Qiu_Wu_Ye_Wang_Liu_2019}:

\begin{align}
f(X_1,X_2,...,X_\mathcal{N}) &= F\left(\sum_{i=1}^{N_1}\Phi(x^{X_1}_i),\sum_{i=1}^{N_2}\Phi(x^{X_2}_i),...,\sum_{i=1}^{N_\mathcal{N}}\Phi(x^{X_\mathcal{N}}_i)\right) \\
    &= F\left(z_1,z_2,...z_\mathcal{N}\right)
\end{align}
where $z_X\in\mathbb{R}^L$ is the latent representation of set $X$ and $\Phi:\mathbb{R}^d\rightarrow\mathbb{R}^L,F:\mathbb{R}^{\mathcal{N}L}\rightarrow\mathbb{R}$ are continuous learned encoder/decoder functions, respectively. The Multiset Deep Sets theorem can be used to describe any observable defined over multiple events.

The EMD is defined over two events $\mathcal{E},\mathcal{G}$, so this reduces to
\begin{equation}
    f(\mathcal{E},\mathcal{G})=F(z_\mathcal{E},z_\mathcal{G}).
\end{equation}
Before operating on this with the $F$ function, we apply the following transformation $T$:

\begin{equation}
\begin{pmatrix}
\Sigma \\
\Delta
\end{pmatrix}
 = 
 \underbrace{
\begin{pmatrix}
1 & 1 \\
1 & -1
\end{pmatrix}}_T
\begin{pmatrix}
z_\mathcal{E} \\
z_\mathcal{G}
\end{pmatrix}
\end{equation}
with $\Sigma,\Delta$ being symmetric, antisymmetric terms respectively. $T$ is invertible so there is no loss of expressivity, while the $\Sigma,\Delta$ basis provides a better inductive bias. This is because under the exchange of $\mathcal{E}\leftrightarrow\mathcal{G}$, $\Sigma\rightarrow\Sigma$ and $\Delta\rightarrow-\Delta$ so $F(\Sigma,\Delta)\rightarrow F(\Sigma,-\Delta)$. These are then averaged and rescaled in \ref{eq:mapfn-symmetric-head}, \ref{eq:mapfn-output} to enforce exchange symmetry, positivity, and identity of indiscernibles. 

\end{document}